\documentclass{sig-alternate-05-2015}
\usepackage{courier}
\usepackage{multirow}
\usepackage{enumerate}
\usepackage{etoolbox}
\usepackage{fixltx2e}
\usepackage{caption}
\usepackage{amsmath}
\usepackage{graphicx}
\usepackage{array}
\newcolumntype{L}[1]{>{\raggedright\let\newline\\\arraybackslash\hspace{0pt}}m{#1}}
\newcolumntype{C}[1]{>{\centering\let\newline\\\arraybackslash\hspace{0pt}}m{#1}}
\newcolumntype{R}[1]{>{\raggedleft\let\newline\\\arraybackslash\hspace{0pt}}m{#1}}
\usepackage{balance}

\begin{document}
\CopyrightYear{2016} 
\setcopyright{acmlicensed}
\conferenceinfo{HT '16,}{July 10 - 13, 2016, Halifax, NS, Canada}
\isbn{978-1-4503-4247-6/16/07}\acmPrice{\$15.00}
\doi{http://dx.doi.org/10.1145/2914586.2914593}
\title{Friendship Maintenance and Prediction in Multiple Social Networks}

\author{
\alignauthor Ka-Wei Roy Lee and Ee-Peng Lim \\
\affaddr{School of Information Systems} \\
\affaddr{Singapore Management University}\\
\email{\{roylee.2013, eplim\}@smu.edu.sg}}

\maketitle
\begin{abstract}
Due to the proliferation of online social networks (OSNs), users find themselves participating in multiple OSNs. These users leave their activity traces as they maintain friendships and interact with other users in these OSNs.  In this work, we analyze how users maintain friendship in multiple OSNs by studying users who have accounts in both Twitter and Instagram. Specifically, we study the similarity of a user's friendship and the evenness of friendship distribution in multiple OSNs. Our study shows that most users in Twitter and Instagram prefer to maintain different friendships in the two OSNs, keeping only a small clique of common friends in across the OSNs. Based upon our empirical study, we conduct link prediction experiments to predict missing friendship links in multiple OSNs using the neighborhood features, neighborhood friendship maintenance features and cross-link features. Our link prediction experiments shows that unsupervised methods can yield good accuracy in predicting links in one OSN using another OSN data and the link prediction accuracy can be further improved using supervised method with friendship maintenance and others measures as features.
\end{abstract}




\section{Introduction}

\textbf{Motivation}. According to the recent Pew Social Media Usage report \cite{pew}, 52\% of online users now use two or more online social networking sites (OSNs). As such, users today may find themselves engaging friends using a number of OSNs. For example, they may ``like'' their friends' posts on Facebook, retweet their friends' tweets on Twitter, and share photos on Instagram. The participation in multiple OSNs implies that users have to stretch and spread their already limited time and attention over the networks, which results in new dynamics in maintenance of friendships. For instance, a user may choose to connect to the same group of friends in multiple OSNs for ease of friendship maintenance, or conversely a user may partition and maintain different groups of friends in different OSNs while keeping only a smaller group of close friends overlapped in multiple OSNs.

This similarity of user's friendship in multiple OSNs also has impact on the evenness of user's friendship in multiple OSNs. For example, a user who maintains high similarity of friendship in multiple OSNs may or may not choose to partition and distribute his friends evenly across multiple OSNs. Our goal in this paper is to investigate the how users maintain friendships in multiple OSNs. Specifically, we study the similarity of users' friendship and the evenness in user's friendship distribution in multiple OSNs.

The study on users' friendship maintenance behavior may provide some new insights to other user behavior studies in multiple OSNs. Lim et al. conducted an empirical study on user's information sharing behavior in six OSNs and found users exhibited varied information sharing behaviors on different OSNs \cite{lim2015}. They postulated that this was due to the difference in user's usage for different OSNs. From friendship maintenance perspective, a possible explanation could be the users were varying their sharing of information to cater for the different groups of ``audience'' (i.e. friends) in different OSNs. Thus, research on friendship maintenance behavior of users can potentially help to provide new insights to other user's behaviors in these OSNs.

The study on friendship in multiple OSNs have real-world applications. In the second part of our study, we extend our empirical research on user's friendship maintenance in multiple OSNs and propose friendship maintenance related features to predict missing links (i.e. friendship) in multiple OSNs. There have been few recent link prediction studies done on \emph{multidimensional networks} which refers to networks with multiple types of links between nodes. Researchers applied neighborhood features such as Common Neighbors and Adamic-Adar on a dimension of network to predict user's links in another dimension within the same network \cite{rossetti}. However, it is important to point out that there are differences between multidimensional networks and in multiple OSNs. For example, the users need to be matched across different networks in multiple OSNs, while users account matching is not required in multidimensional networks. Also, for multiple OSNs, user behaviors in one network are only observed by neighbors in the same network but not the same users's neighbors in another network, while in multidimensional networks, user behaviors are observed by all neighbors of the multidimensional network. As such, the link prediction in our study is different from the previous link prediction studies in multidimensional networks.

\textbf{Research Objectives and Contributions}. This research is conducted on a large real world dataset consisting of about 100,000 users on both Twitter and Instagram with tens of millions online friends. Our research in this paper is divided into two main parts addressing different research questions. In the first part, the research question is how users maintain friendship across networks. We focus on friendship maintenance measures that allow us quantify \emph{friendship overlapping} and \emph{friendship distribution}.  In the second part of our study, we address the research question of how one conducts friendship prediction in the context of multiple social networks. In particular, we would like to explore using the friendship maintenance measures as features to improve the \emph{friendship prediction} accuracy.

\begin{figure}[h]
\centering
  \includegraphics[scale = 0.23]{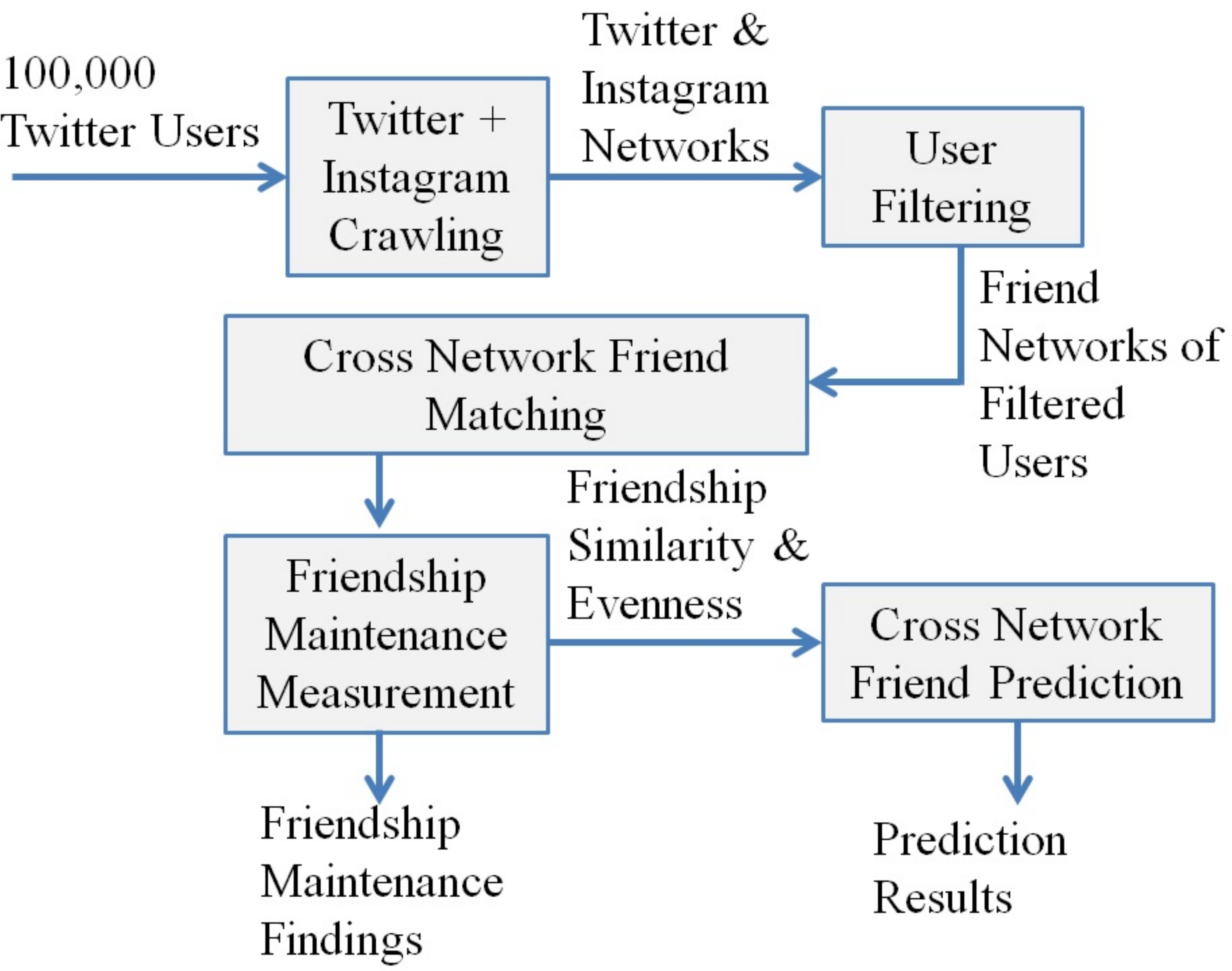}
  \captionof{figure}{Research Framework}
  \label{fig:framework}
\end{figure}

As shown in Figure~\ref{fig:framework}, our proposed research framework begins with data crawling from both Twitter and Instagram to assemble a dataset of base users.  For this set of users, we perform \emph{cross-network friend matching} to identify the Twitter and Instagram friends of the same users.  We then propose several measures for their friendship maintenance behavior.  Finally, we use our findings to design both unsupervised and supervised friend prediction methods.

This work improves the state-of-the-art of social network analysis and link prediction in multiple OSNs. We establish a novel research framework to compare friends in two OSNs. Included in the framework are the measures for evenness of friendship distribution and similarity of friendship across multiple OSNs, as well as the prediction of links in the multiple OSNs settings. The interesting findings derived from our work include:
\begin{itemize}
  \item Most users prefer to maintain roughly the same number of friends in Twitter and Instagram. i.e. evenly distributed friendship across multiple OSNs.
  \item Most users prefer to maintain different friendships in Twitter and Instagram, while keeping only a small clique of common friends across the two OSNs. i.e. low similarity in friendship across multiple OSNs.
  \item Unsupervised methods can yield good accuracy predicting friendship in one network using neighborhood properties of another network. In particular, the Jaccard Coefficient of two users computed in Instagram network can quite accurately predict the link between the two users in Twitter (average F1 score of 0.882).
  \item Supervised method with friendship maintenance measures as features can further improve the accuracy in friendship prediction across multiple OSNs (average F1 score of 0.93).
\end{itemize}

\textbf{Paper Outline}. The rest of the paper is organized as follows. We first describe the construction of our Twitter and Instagram datasets. Next, we propose measures that quantify the evenness of user's friendship distribution and similarity of friendship in multiple OSNs. We then apply the proposed measures to analyse the users' friendship maintenance in Twitter and Instagram networks. Subsequently, we describe the friendship link prediction experiments conducted using friendship features and present the results. Finally, we review related research to this study and conclude this work with possible future research.

\section{Base User Dataset}

In order to study the user friendships in multiple OSNs, we first need to construct a dataset of users who have accounts with both Twitter and Instagram, a popular microblogging site and a photo-sharing social media site respectively. As the two selected OSNs serve different purposes, it is unlikely that the two OSNs cannibalize each other's users. Furthermore, the two OSNs are highly complementary and popular among teen users \cite{pew}. We therefore expect a user on both Twitter and Instagram would generally have the interest to include the same friends in both networks.

We begin by gathering a set of 100,000 Twitter users who have declared their Instagram accounts in their Twitter biography description from \textit{Followerwonk} \footnote{https://moz.com/followerwonk/}, a Twitter analytic platform. Subsequently, the Twitter and Instagram followers and followees of these 100,000 users were crawled using the Twitter and Instagram APIs. However, as some of these Twitter and Instagram accounts have set their privacy settings to ``private'', we are not able to obtain all the followers and followees of the users. We are also only interested in analyzing friendship of average OSN users, thus we further filter away celebrity or popular users who have more than 2,000 followers. At the end, we manage to obtain 97,978 users who have declared both their Twitter and Instagram accounts, and these users constituted the \textbf{base user set}.

\begin{figure}[h]
\centering
\includegraphics[scale = 0.18]{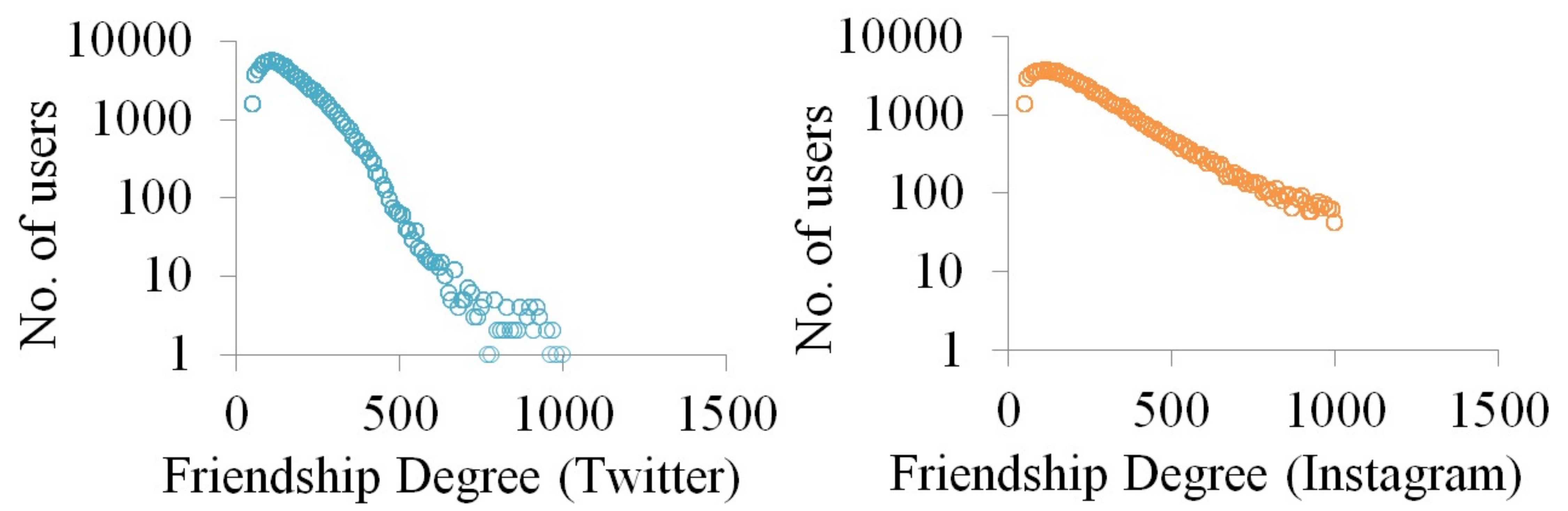}
\caption{Twitter and Instagram Friendship Distribution}
\label{fig:degree}
\end{figure}

Next, we retrieve the Twitter and Instagram friends of the users in \emph{base user set}.  As Twitter and Instagram only capture follower and followee relationships, we define the \emph{friend} of a user to be someone who follows and is followed by the user \cite{xie,java}. An estimated 17 million Twitter friends and 24 million Instagram friends are finally obtained. Figure~\ref{fig:degree} shows the Twitter and Instagram friendship degree distributions. The average Twitter and Instagram friendship degrees for these users are 171 and 245 respectively.

\section{User Friend Matching}
\label{sec:friendmatching}

Before we can study how the users maintain friendships in their Twitter and Instagram accounts, we are required to match the friend accounts in the two OSNs. Unfortunately, very few of the friends have declared both their Twitter and Instagram accounts.  Hence, in this section, we present a few simple but effective ways to match users between OSNs by adapting the methods proposed by Zafarani and Liu \cite{zafarani:connect} and Vosecky et al. \cite{vosecky2009user}, which are quite effective in this context. We match the Twitter and Instagram friends of our base user set using three levels of user matching methods:

\begin{enumerate}

  \item \textbf{Self-Report Matching}.  This method matches the Twitter and Instagram friends of the base user set if these friends declare both their Twitter and Instagram accounts.
  \item \textbf{Username Matching}.  Past research has reported that 59\% of users prefer to use the same username repeatedly on different OSNs for easy recall \cite{zafarani:connect}. Instead of matching all our Twitter and Instagram users by their usernames, we match Twitter users with Instagram users by username when they are the friends of the same user in our base set. This minimizes the possibility of two users being matched because they adopt more popular username.
  \item \textbf{Username Bigram Matching}. Users may tweak their usernames slightly across different OSNs due to the unavailability of their usual usernames. To cater for such situations, we introduce an approximate method which matches the Twitter and Instagram friends of the base users using username bigrams.  Each username is now represented by a vector of bigram weights each of which is the number of occurrences of the bigram in the username.  Cosine similarity is then applied on two username bigram vectors to determine if the two usernames are sufficiently similar.  If the cosine similarity score exceeds a threshold, the two usernames are considered matched.  We adopt a threshold value of 0.63 which is derived by taking the median cosine similarity values of Twitter and Instagram username bigrams of the base users.
\end{enumerate}

\begin{table} [h]
\caption{Number of users and friends matched using different methods}
\label{tab:match}
\centering
\scriptsize
\begin{tabular}{|l|c|c|c|c|}
\hline
Methods            & Self-  &  Username & Username & Total \\
                   & Report &           & Bigram   &       \\ \hline \hline
\# Users Matched   & 17,236 & 1,473,217 & 1,546,645 & 3,037,098 \\ \hline
\# Friends Matched & 22,234 & 1,735,719 & 1,798,457 & 3,556,410 \\ \hline
\end{tabular}
\normalsize
\end{table}

Table~\ref{tab:match} shows the number of friends matched using the above three methods. As expected, the self-report method returns the smallest number of matched friends. A total of 22,234 friends were matched using this method giving an average of $\frac{22,234}{97,978}=0.23$ matched friends per user.  In other words, vast majority of base users do not have their Twitter and Instagram friends matched using self-report. User name matching method, on the other hand, is able to match a total of 1,735,719 friends (in addition to those matched by self-report) or an additional 17.72 friends per user, representing $\frac{17.72}{171}=10.4\%$ and $\frac{17.72}{245}=7.2\%$ of all Twitter and Instagram friends of the base users respectively.  Finally, the username bigram matching method returns yet an additional 1,798,457 matched friends, or 18.36 matched friends per user.  This corresponds to 10.7\% and 7.5\% of all Twitter and Instagram friends respectively. Combining all methods, we are able to match 3,556,410 friends, or 36.3 matched friends per user. Henceforth, we will use all these matched friends in the subsequent analysis.

As there are no ground truth for the validation of the matched friends, we randomly inspected Twitter and Instagram profiles of 100 pairs of matched friend pairs using the username matching and another 100 pairs of matched friends using combined method. We then looked at the visual cues such as their profile photos to assess whether the matching methods are accurate. Among the inspected 100 pairs of matched friends using the exact username matching method, we observed that 77 of the pairs have (i) matching profile photos for their Twitter and Instagram accounts, or (ii) their Twitter profile photos matched with some of the photos posted by the Instagram accounts. Majority of the non-matched friend profiles are due to the users not setting profile picture for their Twitter accounts, thus the actual number of matched pair could be higher than 77. For the username bigram method, 68 of the pairs meet the matching profile photos criteria. This suggests that the user matching methods were able to match the user friends with good accuracy. 

\section{Friendship Maintenance \\ Measurement}
\label{sec:maint}

Before we study how users maintain friendship in Twitter and Instagram, we first propose two measures, \emph{friendship similarity} and \emph{friendship evenness}, to quantify the similarity of user's friendship and the evenness of user's friendship distribution in multiple OSNs respectively.

\subsection{Friendship Similarity}

To ease friendship maintenance, users may choose to overlap their friendships in multiple OSNs. We adapt the \textit{D-Correlation} approach by Berlingerio et. al \cite{berlingerio} to measure this overlap or similarity of friendship across multiple OSNs. D-Correlation was originally designed for multi-dimensional networks where it measures how redundant are two dimensions for existence of a node or an edge.

We use $\mathbb{N}$ to denote a set of OSNs $\{N_1, N_2, \cdots, N_n\}$. We denote the set of friends of a user $x$ in a OSN $N_i$ by $FR(x,N_i)$. We define the friendship similarity of user $x$ among these OSNs, $F_{Sim}(x,\mathbb{N})$, to be the ratio of common friends of $x$ across all OSNs as shown in Equation~\ref{sim_eqn}.

\begin{equation} \label{sim_eqn}
F_{sim}(x,\mathbb{N}) = \frac{|\cap_{N_i \in \mathbb{N}} FR(x,N_i)|}{|\cup_{N_i \in \mathbb{N}} FR(x,N_i)|}
\end{equation}

\begin{figure}[h]
\centering
  \includegraphics[scale = 0.11]{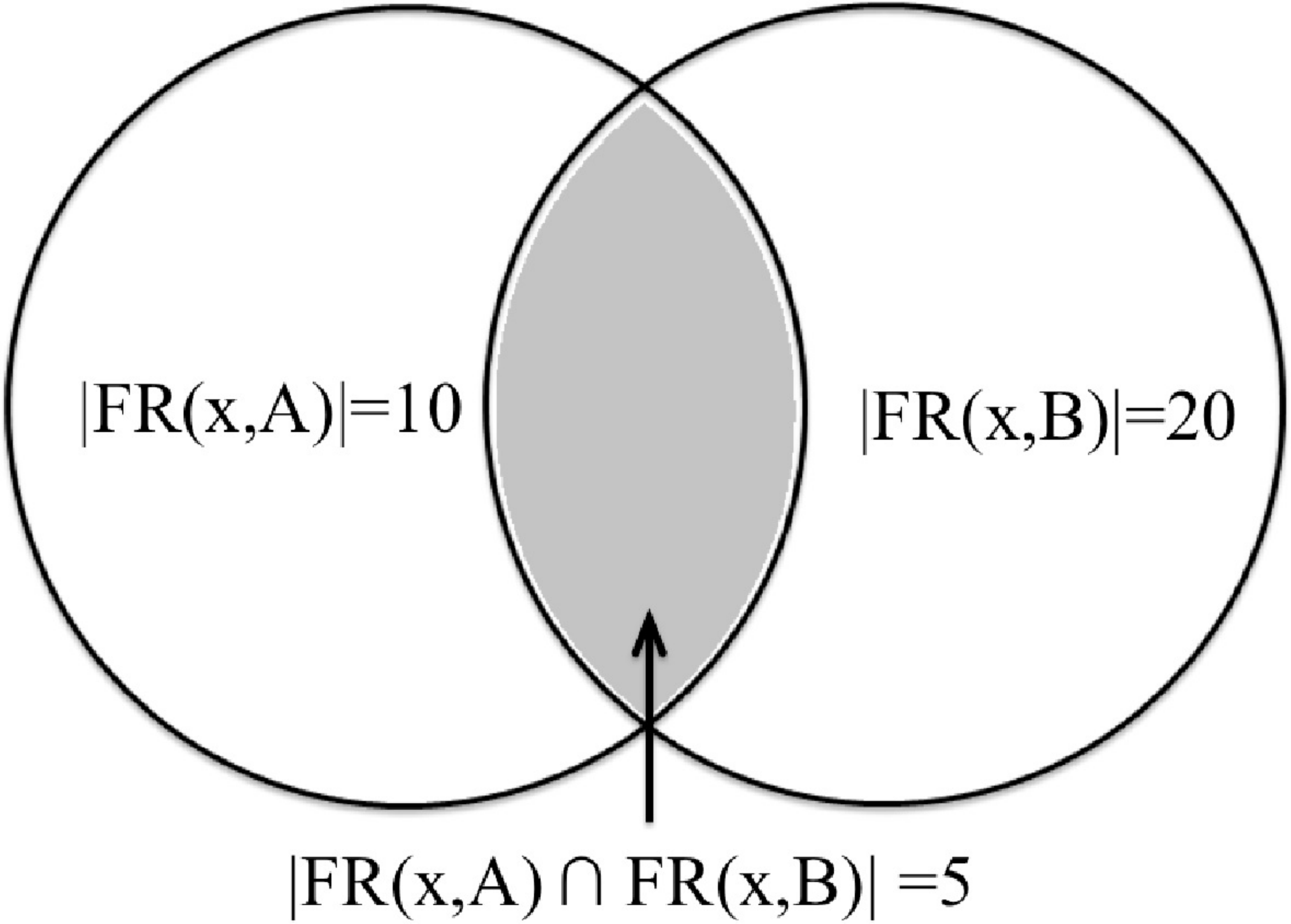}
  \captionof{figure}{Example of user's friendship in two OSNs}
  \label{fig:example}
\end{figure}

\textit{\textbf{Example.}} Figure \ref{fig:example} illustrates the an example of user distributing his friends in two OSNs, \textit{A} and \textit{B}.  The user $x$ have a total of 25 friends; 10 friends in \textit{A}, 20 friends in \textit{B} and 5 of the friends are overlap two OSNs. Thus, the user \textit{x}'s friendship similarity in OSN \textit{A} and \textit{B} will be computed as  $F_{sim}(x,\mathbb{N}) = 5/25 = 0.2$.

\textit{\textbf{Upper Bound of Friendship Similarity.}} The maximum Friendship Similarity value is only achieved when $x$ has the same friends in all OSNs.

The maximum value of for a user's friendship similarity in multiple OSNs is equal to ratio between the minimum and maximum number of friends added to a OSN among the OSNs that the user has participated (as shown in Equation~\ref{maxsim_eqn}). Referencing to the earlier example in Figure~\ref{fig:example}, the maximum possible $F_{sim}$ value for user \textit{x} would be $10/20=0.5$. i.e. user \textit{x} added all his friends in OSN \textit{A} in OSN \textit{B} as well.

\begin{equation} \label{maxsim_eqn}
max(F_{sim}(x,\mathbb{N})) \leq \frac{\min\limits_{N_i \in \mathbb{N}}| FR(x,N_i)|}{\max\limits_{N_i \in \mathbb{N}} |FR(x,N_i)|}
\end{equation}

\subsection{Friendship Evenness}

Suppose that a user \textit{x} divides all his friends among all the $n$ OSNs without overlap, we expect $\frac{1}{n}$ of his friends in each OSN.  Suppose there is a non-zero overlap among his friends across all the OSNs but negligible overlap between subsets of OSNs, and $F_{sim}(x,\mathbb{N})>0$, the \textit{expected ratio of friends x adds to each OSN} is then estimated by $\frac{1}{n}+\frac{F_{sim}(x,\mathbb{N})}{n}$ as shown in Equation~\ref{equal_eqn}.

\begin{equation} \label{equal_eqn}
F_{equal}(x,\mathbb{N}) = \frac{1+(n-1)\cdot F_{sim}(x,\mathbb{N})}{n}
\end{equation}

\noindent
\textbf{Proof.} Suppose $x$ has $N$ unique friends in $\mathbb{N}$. Assume that $x$ distributes her friends evenly across the OSNs.  Let $N_u$ be the number of unique friends in each OSN and let $F$ denote $F_sim(x,\mathbb{N})$. We then expect $x$ to have $N \cdot F$ common friends across the OSNs. In other words, $x$ has $N_u + F\cdot N$ friends in each OSN.  As $N = n \cdot N_u +  F \cdot N$, we obtain $N = \frac{n\cdot N_u}{1-F}$. Each OSN is then expected to have $N_u + F \cdot \frac{n\cdot N_u}{1-F}$ friends in each OSN.  The expected ratio of friends in each OSN is therefore
\begin{equation}
\frac{N_u + F \cdot N}{N} \\
= \frac{N_u + F \cdot \frac{n\cdot N_u}{1-F}}{\frac{n\cdot N_u}{1-F}} \\
= \frac{1+(n-1)\cdot F}{n}
\end{equation}

When $F=0$, the above ratio degenerates to $\frac{1}{n}$ implying that all friends of $x$ are equally divided among OSNs exclusively. When $F=1$, the ratio also becomes $1$ implying that every OSN covers all friends of $x$.  When there are only two OSNs, i.e., $n=2$, the expected ratio of friends in each OSN is $\frac{1+F}{2}$.

However, we would expect that in many circumstances, unevenness exists among the friend counts of the OSNs. For example, a user may maintain a larger group of friends in an OSN $N_i$ while keeping a smaller clique in another network. We thus define the \textit{ratio of friends of a user $x$ in OSN $N_i$ relative to all friends} in Equation~\ref{in_eqn}.

\begin{equation} \label{in_eqn}
F_{in}(x,N_i,\mathbb{N}) = \frac{|FR(x,N_i)|}{|\cup_{N_i \in \mathbb{N}} FR(x,N_i)|}
\end{equation}

Finally, we then define the \textit{evenness of user's friendship distribution} in multiple OSNs as the inverse of summation of difference between the ratio of friends added in each OSN and the expected ratio of friends a user adds to each OSN when the friends are evenly distribution as shown in Equation \ref{even_eqn}.

\begin{equation} \label{even_eqn}
F_{even}(x,\mathbb{N}) = 1 - \sum_{i=1}^{n}\Bigl|F_{in}(x,N_i,\mathbb{N})-F_{equal}(x,\mathbb{N})\Bigr|
\end{equation}

\textit{\textbf{Example.}} Referring to our earlier example in Figure~\ref{fig:example}, $F_{in}(x,A,\{A,B\})$ is $10/25 = 0.4$ and $F_{in}(x,B,\{A,B\})$ is $20/25 = 0.8$. User \textit{x}'s evenness of friendship distribution in OSN \textit{A} and \textit{B} is  $F_{even}(x,\{A,B\}) = 1 - (|0.4-\frac{1+0.2}{2}|+|0.8- \frac{1+0.2}{2}|) = 0.6$.  

Note that $F_{even}(x,\{A,b\})$ measure is also in the range of 0 to 1. Suppose that a user add equal number of friends in the two OSNs with any number of overlap friends among the two OSNs, the user's friendship evenness value will 1. The value for friendship evenness will be 0 is no friend in one of the two networks.

\textit{\textbf{Relationship between Measures.}} There is also an interesting relationship between the upper bound of Friendship Similarity and Friendship Evenness. Based on Equation~\ref{maxsim_eqn}, in order to achieve a maximum friendship similarity value of 1 (i.e., $max(F_{sim}(x,\mathbb{N}))=1$), the minimum and maximum numbers of friends in all the OSNs are identical. That is, user $x$ distributes friendships evenly among all the OSNs ($F_{even}(x,\mathbb{N})=1$). Thus, the more evenly distributed the friends among OSNs, the higher the $max(F_{sim}(x,\mathbb{N}))$.

\section{Empirical Results}

In this section, we apply the friendship similarity and evenness measures to analyze how the 97,978 \textit{base users} maintain their friendships in Twitter and Instagram.

\subsection{Distribution Analysis}

\begin{figure}[h]
\centering
\includegraphics[scale = 0.2]{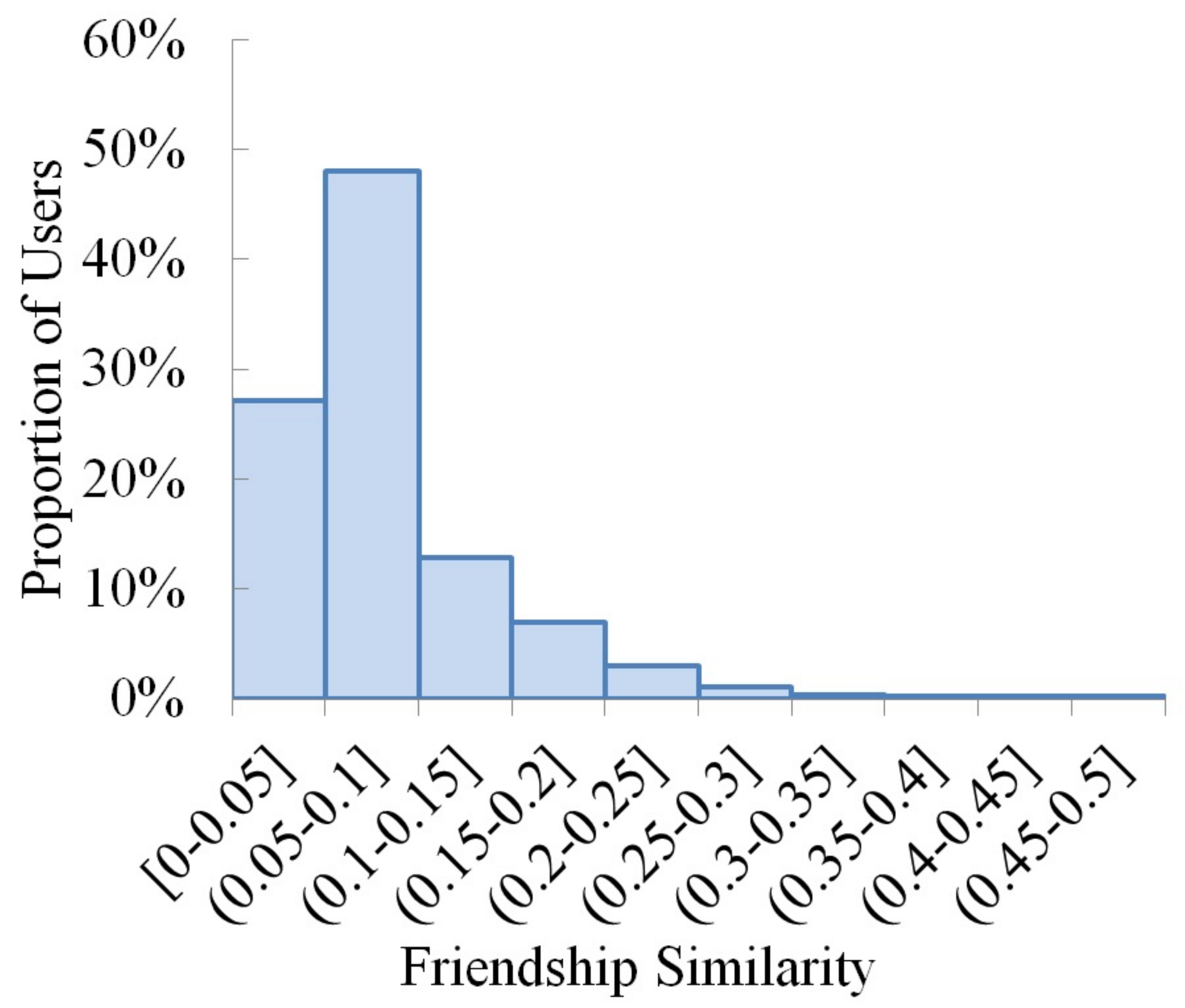}
\caption{Friendship Similarity Distribution}
\label{fig:fsimdist}
\end{figure}

Figure~\ref{fig:fsimdist} shows the distribution of friendship similarity. The average friendship similarity is 0.104. The 1st, 2nd and 3rd quartile friendship similarity values are 0.046, 0.09 and 0.148 respectively.  This left-leaning bell shape distribution suggests that there are very few users who maintained similar friendship in their Twitter and Instagram accounts. Interestingly, this is contrary to our initial hypothesis that user would prefer to have a high friendship similarity for ease of maintenance. There could be a few reasons for the low average friendship similarity; for instance, the users may have maintained low evenness for their friendship in the two OSNs, thus limiting the maximum possible friendship similarity value for the users, or the users simply prefer to maintain different groups of friends in different OSNs.

\begin{figure}[h]
\centering
\includegraphics[scale = 0.2]{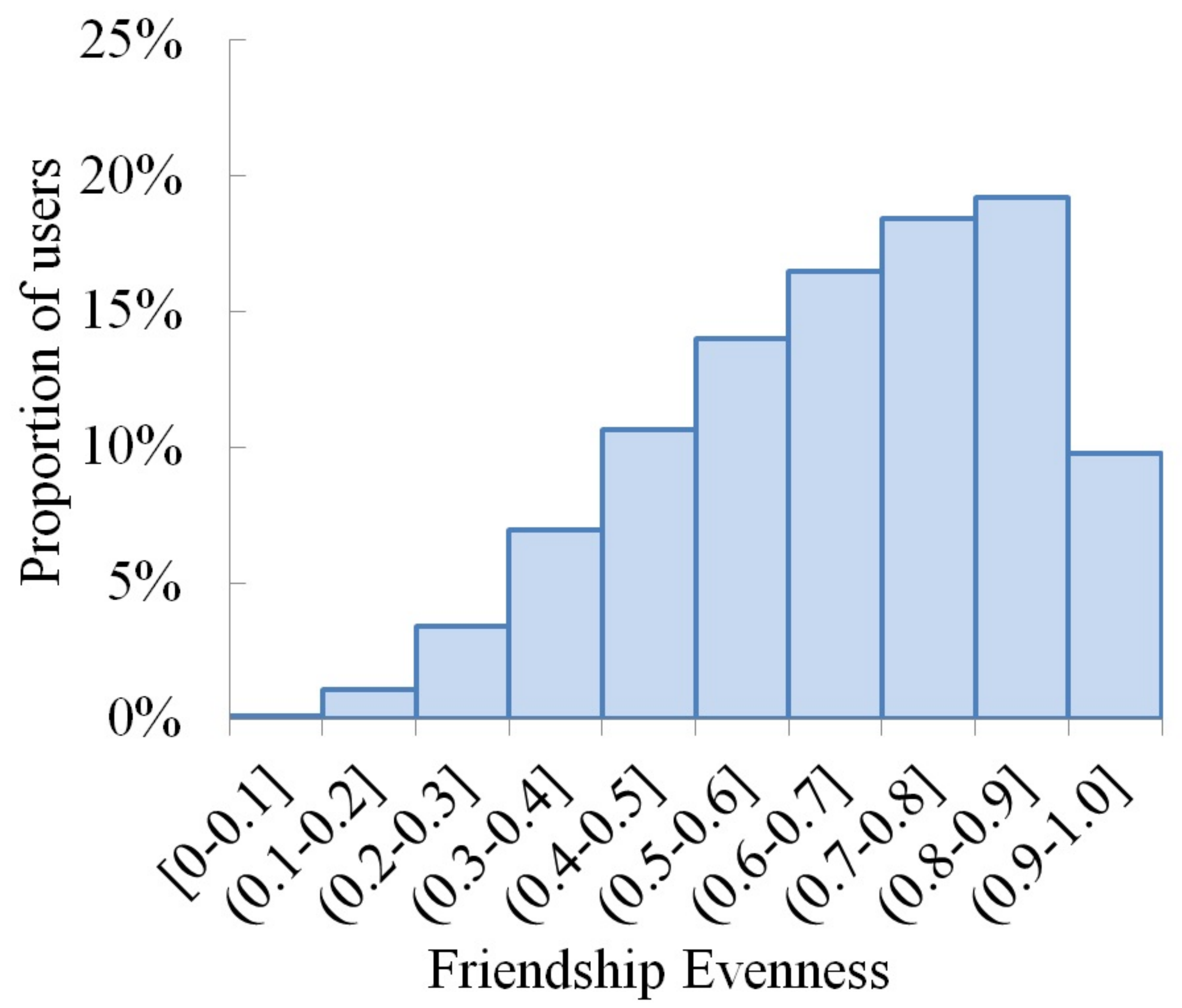}
\caption{Friendship Evenness Distribution}
\label{fig:fevendist}
\end{figure}

Figure~\ref{fig:fevendist} depicts the distribution of friendship evenness of the base users. The average friendship evenness is 0.648, a value much higher than the average friendship similarity.  The 1st, 2nd and 3rd quartile evenness values are 0.534, 0.705 and 0.856 respectively. The distribution is right-leaning, suggesting that most users may prefer to have not overly uneven friendship counts in different OSNs. Also, the right-learning friendship evenness distribution further strengthens our earlier finding that the users tend to prefer to maintain different groups of friends in different OSNs. There could be many reasons for users preference to maintain different friendship in different OSNs. One of the possible reasons could be as suggested by Lim et al. \cite{lim2015}, that users use different OSNs for different purposes or interests, which indirectly motivates the users to connect to different friends in different OSNs. To explain the the user's friendship maintenance behavior, we will study beyond the structural properties of multiple OSNs and investigate the differences in the user interests across different OSNs in our future works.

\subsection{Relationship Between Measures}

\begin{figure}[h]
\centering
\includegraphics[scale = 0.2]{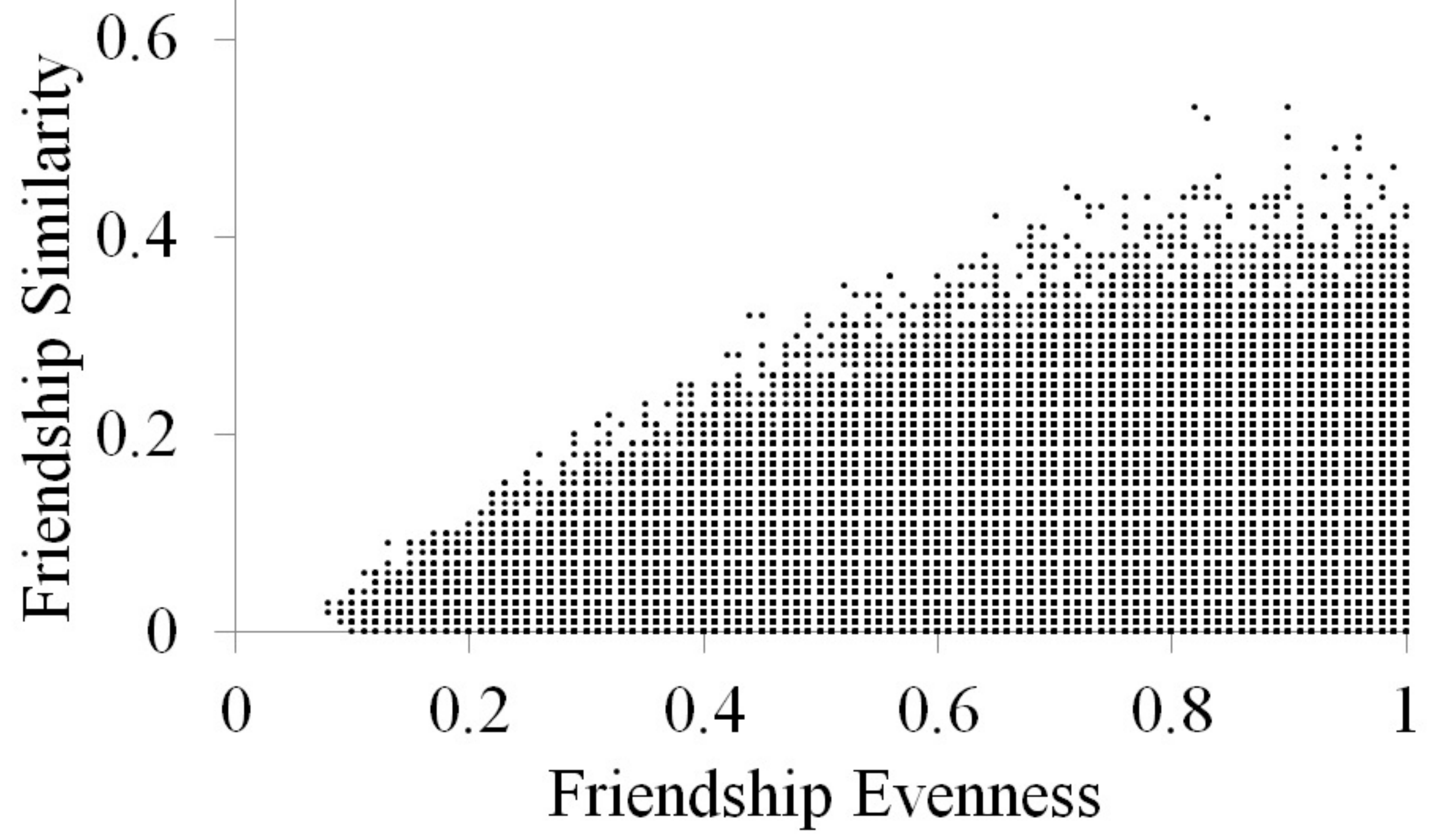}
\caption{Friendship Similarity and Friendship Evenness}
\label{fig:correlation}
\end{figure}

We also examine the relationship between friendship similarity and friendship evenness of users in Figure~\ref{fig:correlation} where each point in the figure represents a user with his friendship similarity and evenness values.

\begin{figure}[h]
\centering
\includegraphics[scale = 0.2]{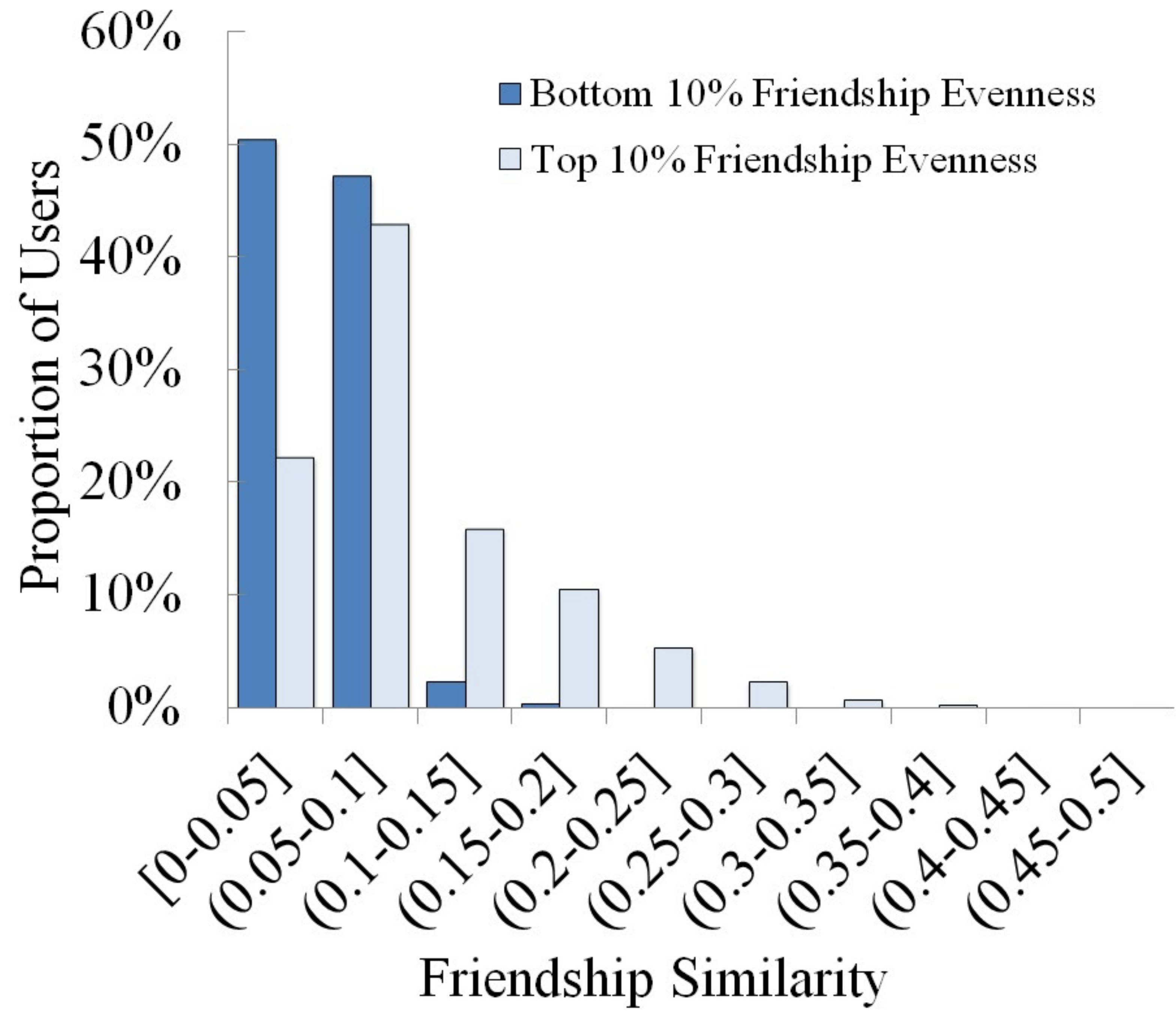}
\caption{Friendship Similarity of Top and Bottom 10\% Friendship Evenness Users}
\label{fig:topbottom}
\end{figure}

Figure~\ref{fig:correlation} shows that as the user's friendship evenness increases, friendship similarity seems to increase its range of values. This supports what we have highlighted in our earlier discussion that the friendship similarity is limited by the friendship evenness. We also further investigate this by showing the friendship similarity distribution of users with top and bottom 10\% friendship evenness in Figure~\ref{fig:topbottom}. The top 10\% friendship evenness users have friendship similarity distribution similar to the overall friendship similarity distribution (as shown in Figure~\ref{fig:fsimdist}), while the bottom 10\%  friendship evenness users have a more left-leaning friendship similarity distribution. The top 10\% friendship evenness users also have an average of friendship similarity of 0.124, slightly higher than the 0.104 friendship similarity of an average user, while the bottom 10\%  friendship evenness users have an average of 0.055 friendship similarity, significantly lower than the average user. However, it is observed that there are quite still a number of users who have high friendship evenness but low friendship similarity.

To investigate the dependency between friendship evenness and similarity, we performed a Chi-squared Test of Independence on the two measures. The test result shows p-value < 2.2e-16, which is lesser than the 0.05 significance level, therefore we reject the null hypothesis that friendship similarity is independent of friendship evenness. The two measures also shows a positive weak correlation of 0.277.


\section{Friendship Link Prediction}
\label{sec:linkprediction}

We now examine how the link prediction in multiple social networks can leverage on the links across networks. Link prediction can come in two forms, namely, prediction of future links and prediction of missing links \cite{nowell,goldberg,taskar}. In our research, we focus on the latter which is useful in applications such as friend recommendations.  As this is the first attempt to conduct link prediction for multiple social networks, we also want to answer the following research questions:
\begin{itemize}
  \item \emph{Can we predict the link between two users in one network using the structural information of the two users in another network?}  Suppose that two users have many common friends in a single OSN, it is likely the they are friends in the OSN. Intuitively, the existence of a link between the two users in one OSN should also increase the likelihood of a link between the users in another OSN.
  \item \emph{Can the friendship maintenance features improve the accuracy of link prediction in multiple online social networks?} Now that we have the friendship similarity and evenness measures, we would like to know if they can make good features for link prediction.
\end{itemize}

\subsection{Task Definitions}

There are two prediction tasks to be performed: (a) \textbf{Twitter Link Prediction (TWLP)} where we predict if two users are friends in Twitter; and (b) \textbf{Instagram Link Prediction (INLP)}, where we predict if two users are friends in Instagram.

We now describe the setup of the training and test data in our the link prediction task. Let $V_{Both}$ be the 97,978 base users who exist in both Twitter and Instagram.  For our base users in Twitter, we define the set of positive instances to be $(u,v)$ pairs such that both $u$ and $v$ are in $V_{Both}$ and $(u,v)$ is an observed link in Twitter. We denote this set of positive instances by $E_{pos}(TWT)$. The set of negative instances, denoted by $E_{neg}(TWT)$, is the set of $(u,v)$ pairs with both $u$ and $v$ from $V_{Both}$ but are not friends in Twitter.  The sets of positive and negative instances for our base users in Instagram are defined in a similar manner.

With the above definitions, we derive 17,651 and 26,241 positive instances for base users in Twitter and Instagram respectively, i.e., $|E_{pos}(TWT)|=17,651$ and $|E_{pos}(INT)|=26,241$.  The numbers look small compared with the size of base users largely because the base users which are selected based on having both Twitter and Instagram accounts do not come from the same user community. Hence, only very few of them know each other on Twitter or Instagram.  In other words, there are many more negative instances making the link prediction tasks highly imbalanced.  Furthermore, there are additional overheads crawling additional data (e.g., friends of neighbors) for each positive and negative instance in the prediction task.  In order to keep the number of instances manageable for the prediction methods, we randomly select 5,000 positive instances and 25,000 negative instances for each run in our prediction tasks. The negative instances are kept to five times that of positive instances. To make the prediction harder, we also check that at least 5,000 negative instances have at least 1 common neighbor in Twitter or Instagram.

\textbf{Unsupervised Prediction task}. For this task, we rank the 5,000 positive and 25,000 negative instances by some ranking measure.  We expect the top ranked instances to be positive if the prediction method works accurately.  In the ideal case, all positive instances are ranked above all negative ones.

\textbf{Supervised Prediction Task}. For this task, we select set of training and test datasets. Each dataset consist of 5,000 positive instances and 25,000 negative instances which are randomly selected. We also check that the instances selected for testing dataset does not exist in the training dataset. We then train a classifier using the training dataset and apply the trained classifier on the test dataset.  This experiment is repeated three times and the results reported are the average of the three runs.

\subsection{Unsupervised Link Prediction Methods}

We propose to use several unsupervised link prediction methods using different \emph{neighborhood features} as ranking measures\cite{newman, adamic}.  These measures involve using the common neighbors between a pair of users $u$ and $v$ to derive some affinity score for ranking the user pair.  These measures are also based on the triadic closure principle in social network analysis \cite{simmel}.  In this work, the following measures are used:
\begin{itemize}
  \item \textbf{Common Neighbors} (\textbf{CN}): This measure counts the number of common neighbors between $u$ and $v$.
  \item \textbf{Jaccard Coefficient} (\textbf{JC}):  This measure returns the fraction of common neighbors between $u$ and $v$.
  \item \textbf{Adamic-Adar} (\textbf{AA}): This measure considers the popularity of common neighbors. The less popular common neighbors are given larger weights as they are added together to derive an affinity score.
\end{itemize}

The above measures are chosen as they were commonly used in link prediction experiments. More formal definitions of them are given at the top of Table~\ref{tab:features}. In Table~\ref{tab:features}, $FR(u,T)$ and $FR(u,I)$ denotes the friends of $u$ in Twitter and Instagram respectively.  While applied to score each of the 5,000 positive and 25,000 negative instance, the measures are computed using all observable link instances in our dataset, i.e., all links excluding those used as positive instances.

There were also recently studies that applied these neighborhood measures in multidimensional networks, where links between users in one dimension are ranked using the neighborhood features of users in another dimension of the same network \cite{rossetti}.  Unlike these existing link prediction works on multidimensional networks, we are now using these neighborhood measures for unsupervised link prediction between users in multiple social networks where users may not have accounts on both networks and users having accounts on both networks may not have their accounts matched.

\textbf{Performance Evaluation.}
We use \emph{F1 at Top K} to evaluate each unsupervised link prediction method.  We first rank all given 30,000 instances by the method's measure in decreasing order.  The \emph{Precision} and \emph{Recall at Top K} are computed by:
\[ Prec@K = \frac{\text{\# correct predictions among top K ranked instances}}{K} \]
\[ Rec@K = \frac{\text{\# correct predictions among top K ranked instances}}{5000} \]
\[ F1@K = \frac{2 \cdot Prec@K \cdot Rec@K}{Prec@K + Rec@K} \]

\begin{figure}[h]
\centering
\includegraphics[scale = 0.22]{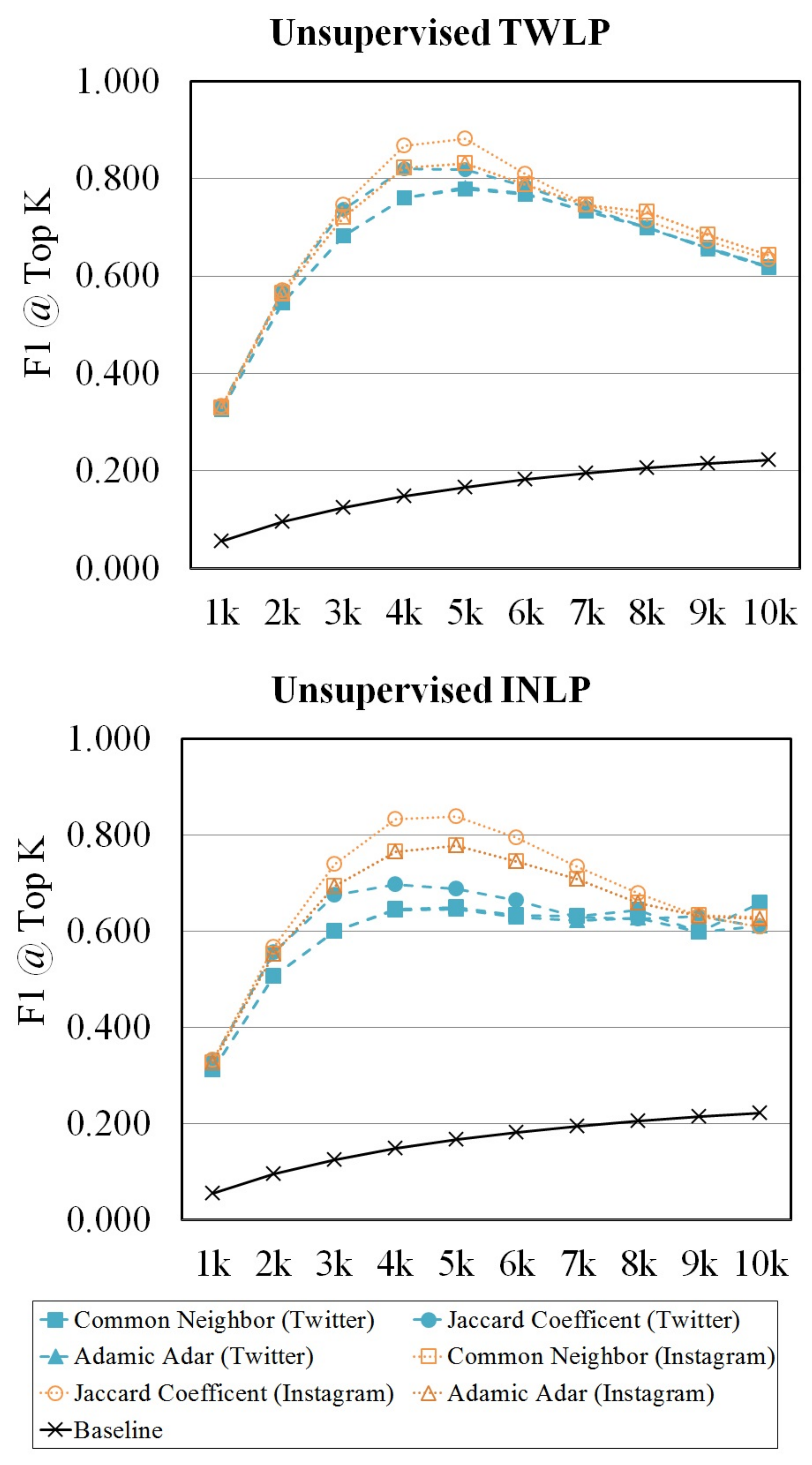}
\caption{F1 scores @ Top K for TWLP and INLP}
\label{fig:f1topk}
\end{figure}

\textbf{Experiment Results.}
Figure~\ref{fig:f1topk} shows F1@K of unsupervised link prediction methods in TWLP and INLP tasks. We introduce a baseline method which returns randomly selected K instances as predicted links. We vary $K$ from 1000 to 10,0000 to examine the performance of each method.

As shown in the figure, all the unsupervised methods perform significantly (3 to 4 folds) better than the random baseline in both TWLP and INLP tasks. While the baseline method increases gradually with larger K values due to increasing recall, most of the other methods improve their F1@K only up K=4000 or K=5,000.  Beyond which, their F1@K drops. This is because these methods are able to rank positive instances more highly than negative instances.

Interestingly, the figure also shows that the prediction methods using Instagram links outperform those using Twitter links even when the prediction task involves Twitter link prediction, i.e., TWLP.  In particular, the method using Jaccard Coefficent on Instagram links (i.e., $\textbf{JC}_I$ ) outperforms the rest for almost all K values, achieving the highest F1 scores of 0.882 and 0.838 for TWLP and INLP tasks respectively for top 5,000 ranked results. A possible explanation of the above findings could be that the users have higher friendship degrees in Instagram than Twitter. Two users who are friends in Twitter are likely to have common friends in Instagram. Even though the Twitter neighborhood measures performed worse than Instagram neighborhood measures, they still yield good results (up to 0.689 for F1@5K) in predicting links between users in Instagram. This suggests that predicting links in one OSN using the neighborhood information of another OSN can yield very respectable accuracy.

\subsection{Supervised Link Prediction Methods}

For supervised link prediction, we use Support Vector Machine (SVM) with linear kernel as the binary classifier trained with each instance represented as a feature vector.  SVM is chosen because of its relatively good results in other link prediction tasks.  We also consider three types of features as shown in Table~\ref{tab:features}.  The \textbf{neighborhood features} are the scores from different measures used in unsupervised link prediction methods.  By including the neighborhood features, the supervised methods can hopefully achieve at least the good accuracy of the unsupervised methods.

We introduce a binary \textbf{cross network feature} \textbf{CL} which returns 1 if the users of the instance are friends in another network, and 0 otherwise.  For example, in the case of TWLP task, a $(u,v)$ instance is assigned a CL feature value of 1 if and only if $u$ and $v$ are friends in Instagram. This feature is included because having a friendship in another OSN should increase the odd of the users having friendship in the target OSN.

\begin{table}[t]
\centering
\caption{Link Prediction Features}
\label{tab:features}
\begin{tabular}{|l|l|}
\hline
\textbf{Feature} &  \textbf{Description}  \\ \hline \hline
\multicolumn{2}{|l|}{Neighborhood features} \\ \hline
$\textbf{CN}_T$  & $|FR(u,T) \cap FR(v,T)|$  \\
$\textbf{JC}_T$  & $\frac{|FR(u,T) \cap FR(v,T)|}{|FR(u,T) \cup FR(v,T)|}$  \\
$\textbf{AA}_T$  & $\sum_{z \in FR(u,T) \cap FR(v,T)} \frac{1}{log |FR(z,T)|}$  \\
$\textbf{CN}_I$  & $|FR(u,T) \cap FR(v,I)|$  \\
$\textbf{JC}_I$  & $\frac{|FR(u,I) \cap FR(v,I)|}{|FR(u,I) \cup FR(v,I)|}$  \\
$\textbf{AA}_I$  & $\sum_{z \in FR(u,I) \cap FR(v,I)} \frac{1}{log |FR(z,I)|}$  \\ \hline \hline
\multicolumn{2}{|l|}{Common Neighbor Friendship Maintenance features} \\ \hline
$\textbf{HEHS}_T$ & $\frac{|\{z \in FR(u,T) \cap FR(v,T) | F_{sim}(z) \mbox{ is high}, F_{even}(z) \mbox{ is high}\}|}{|FR(u,T) \cup FR(v,T)|}$ \\
$\textbf{HELS}_T$ & $\frac{|\{z \in FR(u,T) \cap FR(v,T) | F_{sim}(z) \mbox{ is low}, F_{even}(z) \mbox{ is high}\}|}{|FR(u,T) \cup FR(v,T)|}$ \\
$\textbf{LEHS}_T$ & $\frac{|\{z \in FR(u,T) \cap FR(v,T) | F_{sim}(z) \mbox{ is low}, F_{even}(z) \mbox{ is low}\}|}{|FR(u,T) \cup FR(v,T)|}$ \\
$\textbf{LELS}_T$ & $\frac{|\{z \in FR(u,T) \cap FR(v,T) | F_{sim}(z) \mbox{ is high}, F_{even}(z) \mbox{ is low}\}|}{|FR(u,T) \cup FR(v,T)|}$ \\
$\textbf{HEHS}_I$ & $\frac{|\{z \in FR(u,I) \cap FR(v,I) | F_{sim}(z) \mbox{ is high}, F_{even}(z) \mbox{ is high}\}|}{|FR(u,I) \cup FR(v,I)|}$ \\
$\textbf{HELS}_I$ & $\frac{|\{z \in FR(u,I) \cap FR(v,I) | F_{sim}(z) \mbox{ is low}, F_{even}(z) \mbox{ is high}\}|}{|FR(u,I) \cup FR(v,I)|}$ \\
$\textbf{LEHS}_I$ & $\frac{|\{z \in FR(u,I) \cap FR(v,I) | F_{sim}(z) \mbox{ is high}, F_{even}(z) \mbox{ is low}\}|}{|FR(u,I) \cup FR(v,I)|}$ \\
$\textbf{LELS}_I$ & $\frac{|\{z \in FR(u,I) \cap FR(v,I) | F_{sim}(z) \mbox{ is high}, F_{even}(z) \mbox{ is low}\}|}{|FR(u,I) \cup FR(v,I)|}$ \\ \hline \hline
\multicolumn{2}{|l|}{Cross Network features} \\ \hline
\textbf{CL} & $
               \begin{cases}
                      1  & \mbox{if } (u,v) \mbox{ exists in another network}\\
                      0 & \mbox{otherwise}
               \end{cases}
              $ \\
\hline
\end{tabular}
\end{table}

Finally, we also include a group of features known as \textbf{common neighbor friendship maintenance features}. While the neighborhood features in one OSN yield reasonable or even good results in unsupervised link prediction in another OSN, the features may not work very well when the common neighbors demonstrate friendship maintenance behavior that prevent friendship inference across OSNs. For example, a common neighbor between users $u$ and $v$ in Instagram who maintain separate friends in Twitter and Instagram does not increase the likelihood of friendship between $u$ and $v$ in Twitter.  The common neighbor friendship maintenance features are obtained by dividing all common neighbors who are present in both Twitter and Instagram into four different categories: namely: (a) high friendship evenness and high friendship similarity; (b) low friendship evenness and high friendship similarity; (c) high friendship evenness and low friendship similarity; and (d) low friendship evenness and low friendship similarity.  We say that a user has high (or low) friendship evenness if her friendship evenness is greater than (or not greater than) the average friendship evenness value.  We define the user with high or low friendship similarity in the same way.  These common neighbor friendship maintenance features are shown in Table~\ref{tab:features}.

We use six different feature configurations in our supervised link prediction methods as follows:
\begin{itemize}
  \item \textbf{NBO}: Neighborhood features only
  \item \textbf{NFM}: Common Neighbor Friendship Maintenance features only
  \item \textbf{NBOFM}: Neighborhood and Common Neighbor Friendship Maintenance features
  \item \textbf{NBCL}: Neighborhood and Cross Network features
  \item \textbf{NFMCL}: Common Neighbor Friendship Maintenance and Cross Network features
  \item \textbf{ALL}: All features
\end{itemize}

\textbf{Performance Evaluation.}
We conduct three runs of TWLP and INLP experiments and report the average precision, recall and F1 score of each method.  For each run, we use a sample of 5,000 user pairs with friendship and 25,000 user pairs without friendship as the positive and negative instances respectively for training a SVM classifier, and another sample of 5,000 user pairs with friendship and 25,000 user pairs without friendships for testing.  We conducted altogether three runs of training and test evaluation.

\begin{table} [h]
\caption{Link Prediction Results by Supervised Methods}
\label{tab:svmresult}
\centering{
\begin{tabular}{|l|l|c|c|c|}
\hline
Tasks            & Methods& Avg Prec.       & Avg Recall       & Avg F1               \\
\hline \hline
\multirow{5}{*}{TWLP} & \textbf{NBO}    & 0.954           & 0.873            & 0.911           \\ \cline{2-5}
                      & \textbf{NFM}    & 0.955           & 0.830	        & 0.888             \\ \cline{2-5}
                      & \textbf{NBOFM}  & 0.953           & 0.875            & 0.912             \\ \cline{2-5}
			          & \textbf{NBCL}   & 0.976           & 0.887            & 0.929             \\ \cline{2-5}
			          & \textbf{NFMCL}  & \textbf{0.979}  & 0.861            & 0.916             \\ \cline{2-5}
                      & \textbf{ALL}    & 0.973           & \textbf{0.891}   & \textbf{0.930}   \\ \cline{2-5} \cline{2-5}
                      & $\textbf{JC}_I$ & 0.882           & 0.882            & 0.882             \\ \hline \hline
\multirow{5}{*}{INLP} & \textbf{NBO}    & 0.942           & 0.832            & 0.883  \\ \cline{2-5}
                      & \textbf{NFM}    & 0.959           & 0.721	        & 0.823            \\ \cline{2-5}
                      & \textbf{NBOFM}  & 0.942           & 0.833            & 0.884            \\ \cline{2-5}
                      & \textbf{NBCL}   & 0.958           & 0.838            & 0.894            \\ \cline{2-5}
                      & \textbf{NFMCL}  & \textbf{0.971}  & 0.74             & 0.84             \\ \cline{2-5}
			          & \textbf{ALL}    & 0.956           & \textbf{0.841}   & \textbf{0.895}   \\ \cline{2-5} \cline{2-5}
                      & $\textbf{JC}_I$ & 0.838           & 0.838            & 0.838             \\ \hline \hline
\end{tabular}
}
\end{table}

\textbf{Experiment Result.}
Table~\ref{tab:svmresult} shows the results of supervised link prediction for TWLP and INLP tasks.  In these experiments, all the feature configurations yield better precision than recall. Most of them have F1 higher than the best F1 scores of the unsupervised methods (i.e., $\textbf{JC}_I$ ). Generally, according to F1, the configuration using all features outperforms other methods. Although the Common Neighbor Friendship Maintenance (\textbf{NFM}) features  performed slightly worse than the Neighborhood (\textbf{NBO}) features, the \textbf{NFM} features still managed to achieve a reasonably good F1 score of 0.888 and 0.823 for TWLP and INLP tasks respectively. This suggests that we are able to predict, with reasonable accuracy, the friendship between users using the common neighbor's friendship maintenance behavior as features. The addition of Cross Network (\textbf{CL}) feature also improves the results of \textbf{NFM} and \textbf{NBO} features. Interestingly, the configuration with Common Neighbor Friendship Maintenance and Cross Network features (i.e., \textbf{NFMCL}) yield the best precision result in both TWLP and INLP task. This suggests that the existence of a link between the two users in one OSN increases the likelihood of a link between the users in another OSN.

A possible reason for Common Neighbor Friendship Maintenance (\textbf{NFM}) features performing slightly worse than the Neighborhood (\textbf{NBO}) features could be due to the lack of common neighbors with friendship maintenance measures who are also base users. Thus we re-examined the supervised link prediction results and determined the accuracy of link prediction for test instances that have at least one common neighbor who is also a base user.

\begin{table} [h]
\caption{Link Prediction Results of Test Instances with at Least 1 Base User Common neighbor}
\label{tab:svmsubset}
\centering{
\begin{tabular}{|l|l|c|c|c|}
\hline
Task           & Methods& Avg Prec.       & Avg Recall       & Avg F1               \\
\hline \hline
\multirow{2}{*}{TWLP} & \textbf{NBO}    & 0.948           & 0.970            & 0.959           \\ \cline{2-5}
                      & \textbf{NFM}    & \textbf{0.971}           & \textbf{0.994}	        & \textbf{0.982}             \\ \cline{2-5}            \hline \hline
\multirow{2}{*}{INLP} & \textbf{NBO}    & 0.938           & 0.959            & 0.949  \\ \cline{2-5}
                      & \textbf{NFM}    & \textbf{0.976}           & \textbf{0.999}	        & \textbf{0.987}            \\ \cline{2-5}              \hline \hline
\end{tabular}
}
\end{table}

As shown in Table~\ref{tab:svmsubset}, our \textbf{NFM} features only method outperformed the method using \textbf{NBO} features by precision, recall and F1 score in both TWLP and INLP tasks.  This suggests that there were several occasions where the \textbf{NBO} features only method wrongly labeled a positive instance as negative but these instances are correctly labeled by \textbf{NFM} features.

Upon further examination of these test instances, we found that although each user pair have very few common neighbors, the common neighbors actually falls in the \textit{low friendship evenness and high friendship similarity} friendship maintenance category (i.e., LEHS). The users in LEHS connect to more friends in either Twitter or Instagram, while keeping a smaller and potentially closer clique of common friends across the two OSNs. Thus, a pair of users with a LEHS common neighbor are more likely to be friends especially when they belong to the smaller clique of friends in one of the OSNs.

\section{Related Works}
\label{sec:related}

In this section, we review thee groups of existing research works related to our research. The first group is the research studies on structural properties and user behaviors in multiple OSNs. The second group discusses link prediction conducted in multidimensional networks. Finally, the last group focuses on triadic closure property in OSNs, which is often used in link prediction.

The study on structural properties and user behaviors in multiple OSNs is an emerging topic and the research subject has been gaining attractions in recent years. Magnani and Rossi \cite{magnani} did a study on the structural properties in multiple OSNs and proposed to represent multiple OSNs as a \emph{multi-layer network}. They had also extended the degree and closeness centrality measures to multi-layer network. Their work however did not consider other network structural properties or behaviors such as the friendship similarity and evenness across networks. The linkage of user accounts across multiple OSNs belong to the same person is also a widely studied topic \cite{zafarani:connect, zhang2014}. With wider adoption of the new user linkage methods by proposed by previous research works, researchers also studied user behaviors across multiple OSNs.  Benevenuto, et. al, performed a macro-level analysis of user behaviors such as browsing and content posting at different OSNs \cite{benevenuto}. Zafarani and Liu conducted an empirical study on users in 20 social media sites and showed that the most users join and stay active in less than 3 social media sites \cite{zafarani:join}. Kumar et al. analyzed the user migration patterns across seven OSNs \cite{kumar2011}.

Unlike the existing works on user behaviors across multiple OSNs, our study focuses on the friendship maintenance behavior of users when they join multiple OSNs. Our study analyzes if a user would prefer to add a friend in multiple OSNs or simply maintain and restrict the friend to a particular OSN only. The findings of our work provide new perspectives to the existing studies on user behaviors in multiple OSNs. For instance, Ottoni et al. studied the users' activities across Twitter and Pinterest and found that the user usage patterns across the two OSNs differ significantly \cite{ottoni2014}. They found that users tend to post items to Pinterest before posting them on Twitter. Using the insights from our studies, a possible explanation for the observed user behaviors in Ottoni et al's study could be due to the low user friendship similarity across the multiple OSNs and the users were maintaining different groups of friends in different OSNs, thus there was a need for users to re-post the content on multiple OSNs so as to disseminate the information to all friends in different OSNs. Similar explanation could also be made for the study conducted by Lim et al. where they found that users exhibited varied information sharing behaviors on different OSNs \cite{lim2015}; the users, who may maintained low friendship similarity, were catering for the different group of ``audience'' (i.e. friends) in different OSNs. Future works could be done to investigate the impact of friendship maintenance on other user behaviors such as information adoption and diffusion.

There were few link prediction studies done on multidimensional networks. Rossetti et. al performed supervised and unsupervised multidimensional link predictions on the DBLP and IMDb networks \cite{rossetti}. In that study, the researchers used neighborhood features such as Common Neighbors and Adamic-Adar to predict user collaboration in the different dimensions of a network. For example, they predicted the collaboration of authors in DBLP with the publishing venues defined as the dimensions. Our link prediction experiment differs from the previous study as we predict friendship of users in different OSNs instead of different dimensions of the same network. Multiple OSNs is quite different from multidimensional networks as there are unmatched user accounts across multiple OSNs while user accounts matching is not required in multidimensional OSN. Furthermore, our friendship link prediction methods not only consider friendship neighborhood features but also friendship maintenance features.

Another related field of work is the study on triadic closure property in social networks. The triadic closure property been widely studied for many years even before the rise of OSNs \cite{simmel,wasserman}. In recent years, researchers modeled and studied the process of triadic closure in OSNs. For example, Romero and Kleinberg had empirically investigated the triadic closure process in Twitter network \cite{romero}. Lou, et. al, performed prediction of reciprocal relationships and triadic closure process in Twitter. They also developed a model to accurately predict 90\% of the reciprocal relationships in Twitter and to predict the triadic closure process among users \cite{lou}. Our study builds on the existing works and focus on how similarity and evenness of friendship across OSNs affect the likelihood of the triadic closure.

\section{Conclusion and Future Works}
\label{sec:conclusion}

In this paper, we studied how users manage and maintain friendships across multiple social networks. We constructed a base set of about 100,000 users with Twitter and Instagram accounts and studied the friendship of these users in the two OSNs. We introduced friendship similarity to measure the similarity of friendships between two OSNs. A friendship evenness measure was also defined to quantify the degree of balance a user maintains for the number of friendships in different OSNs. We shown that most users prefer to maintain different friendships in different OSNs, while keeping only a small clique of common friends across OSNs.

We also investigated link prediction in multiple OSNs using unsupervised and supervised methods. We shown that the conventional unsupervised methods using neighborhood features perform well even when we predicted links in one OSN using only the network structural properties from another OSN. We also proposed a set of network features and applied them to supervised link prediction method. The experiments shown that the supervised methods with suitable feature sets improved the accuracy over that of unsupervised methods.

To conclude, we note that this research is among the very few conducted on multiple social networks. While we have shown that the concepts of friendship similarity and evenness are important, they need to be generalized beyond just two OSNs. As part of the future work, we plan to expand the study to include larger and more diverse OSNs with overlapping user communities. The content generated by users can be further studied so as to provide more insights about the way users manage the different OSNs.

\section{Acknowledgements}
This work is supported by the National Research Foundation under its International Research Centre@Singapore Funding Initiative and administered by the IDM Programme Office, and National Research Foundation (NRF).

\balance

\bibliographystyle{plain}
\bibliography{ref}

\end{document}